\documentclass[prd,twocolumn]{revtex4}

\usepackage{graphicx}
\usepackage{amsmath,amssymb,amsfonts}
\usepackage{hyperref}

\begin{document}

\title{
Energy Conservation and Cosmological Redshift
from the Brown--York Quasi-Local Energy
}

\author{Bjoern S. Schmekel}
\affiliation{Department of Physics, College of Studies for Foreign Diploma Recipients at the University of Hamburg, 20355 Hamburg, Germany}

\email{bss28@cornell.edu}

\begin{abstract}
The question whether energy is conserved in an expanding universe
remains one of the central conceptual issues of relativistic
cosmology. The absence of a global timelike Killing vector and the
lack of a local gravitational energy density often lead to the
conclusion that cosmological redshift represents a genuine violation
of energy conservation.
In this work we reconsider this problem within the framework of
Brown-York quasi-local energy and its associated balance law.
For comoving spherical boundaries in FLRW spacetimes we show that
the ordinary stress-energy flux through the boundary vanishes
identically for arbitrary perfect fluids, while the unreferenced
Brown-York energy evolves entirely due to a purely geometrical
boundary contribution.
We further demonstrate that the timelike boundary world tube
possesses a natural dynamical reference geometry.
The corresponding reference subtraction removes the universal
background contribution of the unreferenced Brown-York energy and
leads to a quasi-local energy whose leading nontrivial term exactly
reproduces the enclosed matter energy.
For radiation dominated cosmologies, the resulting expansion is
$E_{\rm BY} = H^2R^3 /2 - H^4R^5 / 8 + \mathcal{O}(R^7)$, 
where the leading term coincides precisely with the energy of the
enclosed radiation responsible for cosmological redshift.
Higher-order contributions naturally possess the structure expected
for gravitational self-energy corrections.
These results suggest that cosmological redshift can be understood
within a fully quasi-local conservation framework.
The apparent loss of photon energy therefore need not indicate a
breakdown of energy conservation but may instead reflect the
fundamentally quasi-local character of gravitational energy in
general relativity.
\end{abstract}
\maketitle

\section{Introduction}

Is energy conserved in an expanding universe?

The question has accompanied relativistic cosmology since its
earliest days. Photons continuously lose energy due to cosmological
redshift and the total matter energy inside a comoving volume
generally changes with time. Consequently, it is frequently stated
that energy is not conserved in an expanding universe.

Such conclusions, however, depend crucially on the notion of energy
being employed.
Generic cosmological spacetimes admit no timelike Killing vector and
therefore possess no globally conserved Noether energy.
Moreover, general relativity does not admit a unique local
gravitational energy density due to the equivalence principle.

These observations do not necessarily imply a breakdown of energy
conservation. Rather, they may indicate that both local and global
notions of energy are inadequate in gravitating systems and should be
replaced by quasi-local concepts associated with finite regions and
their boundaries.

Among the various quasi-local constructions proposed in the
literature, the Brown-York quasi-local energy (QLE)
\cite{Brown:1992br} occupies a distinguished position because it
follows directly from the Hamilton-Jacobi analysis of the
Einstein-Hilbert action supplemented by the York boundary term
\cite{PhysRevLett.28.1082}. The formalism naturally provides
quasi-local surface densities of energy, momentum and stress together
with an exact balance law.

An important feature of the Brown-York construction is the presence
of a reference action

\begin{equation}
S_0=S_0[\gamma_{ij}],
\label{S0}
\end{equation}

which fixes the zero point of the quasi-local energy.
The original formalism does not uniquely determine this functional.
It merely requires that it depends on the induced metric of the
timelike boundary world tube.

In many applications, a static Euclidean embedding has effectively
become the canonical choice.
While this prescription successfully reproduces the ADM mass in
asymptotically flat spacetimes, its application to cosmology leads to
well-known difficulties. In particular, the Brown--York energy of a
spatially flat FLRW universe may vanish identically despite the
presence of nonvanishing matter and radiation content.

This observation raises a natural question.
Since the reference term is fundamentally a functional of the entire
boundary metric rather than of an individual spatial cross section,
should the reference geometry itself not inherit the dynamical nature
of the boundary world tube?

The purpose of the present work is to investigate this possibility.

We first evaluate the Brown--York conservation law for comoving
boundaries in FLRW spacetimes and show that ordinary matter fluxes
through the boundary vanish identically.
The evolution of the unreferenced Brown--York energy is therefore
entirely governed by a purely geometrical boundary term, which admits
a direct interpretation as work performed by Brown--York surface
stresses.

We then consider a dynamical reference geometry adapted to the
intrinsic geometry of the timelike boundary world tube.
The resulting reference subtraction removes the universal background
contribution of the unreferenced Brown--York energy and leads to a
remarkably simple small-sphere expansion,

\begin{equation}
E_{\rm BY} = \frac{H^2R^3}{2} - \frac{H^4R^5}{8} + \mathcal{O}(R^7).
\end{equation}

The leading term reproduces exactly the matter energy enclosed by the
boundary,

\begin{equation}
\frac{H^2R^3}{2} = \frac{4}{3} \pi \rho R^3,
\end{equation}

while the higher-order contributions naturally possess the structure
expected for gravitational self-energy corrections.

For radiation dominated cosmologies, the leading contribution
precisely reproduces the energy responsible for cosmological
redshift.
This suggests that the apparent loss of photon energy may be
understood within a fully quasi-local conservation framework and does
not necessarily imply a breakdown of energy conservation.

More generally, the present results indicate that the dynamical
geometry of the boundary itself may provide important guidance in the
selection of physically meaningful reference terms within the
Brown--York formalism.


\section{Brown-York Quasi-Local Energy}

Briefly, the construction proceeds by varying the Einstein-Hilbert
action, supplemented by the York boundary term
\cite{PhysRevLett.28.1082}, with respect to the induced metric
$\gamma_{ij}$ on a timelike boundary ${}^3B$. The resulting boundary
term is built from the extrinsic curvature $\Theta_{ij}$ of ${}^3B$
(cf. Table~\ref{notation}), which measures how ${}^3B$ curves within
the four-dimensional manifold $M$. Projecting the resulting surface
stress-energy-momentum tensor $\tau_1^{ij}$ onto the future-directed
timelike unit vector $u^i$, normal to $B$ within ${}^3B$, and
integrating over a spatial two-surface
$B={}^3B\cap\Sigma_t$ then yields the quasi-local energy introduced
below. For a pedagogical review of this and related quasi-local
constructions, see, e.g., Ref.~\cite{Szabados:2004vb}.

Unless stated otherwise, all quantities in the following
refer to the unreferenced Brown--York construction and are denoted by the sub- or superscript ''1'',
i.e. no reference subtraction is assumed at this stage.
The balance law itself is independent of the choice of
reference action and therefore provides the natural starting
point of our analysis.
The Brown--York surface stress-energy-momentum tensor is

\begin{equation}
\tau_{1}^{ij} = -\frac{1}{\kappa} \left( \Theta\gamma^{ij} - \Theta^{ij} \right),
\end{equation}

where $\kappa=8\pi$ in geometrized units. 
The Brown--York energy enclosed by a two-surface $B$ is

\begin{equation}
E_1 = \int_B d^2x \sqrt{\sigma} u_i u_j \tau_1^{ij}.
\end{equation}

An important property of this energy is the exact balance law which is the closest to a conservation law which is known \cite{Schmekel:2018wbl,AndyPrivate}

\begin{figure}
\scalebox{0.3}{\includegraphics{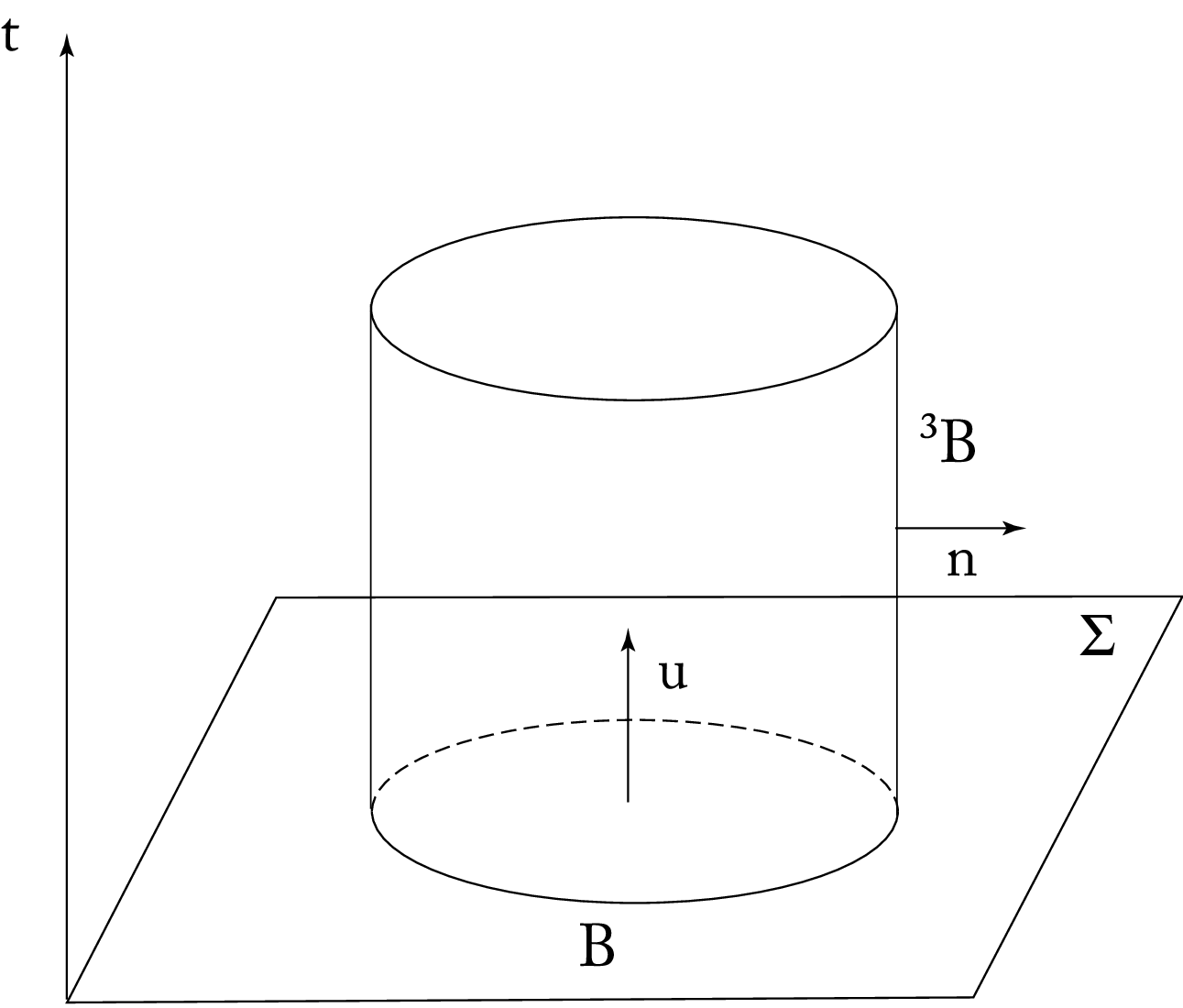}}
\caption{The time evolution of $B$ is $^3 B$. Their unit normal is $n^\mu$ in both cases. The unit normal of a time-slice
$\Sigma$ embedded in the four-dimensional manifold $M$ is denoted by $u^\mu$. Note that one dimension has been suppressed,
i.e. the boundary $B$ which is shown as a one-dimensional line represents a two-dimensional boundary.
Cf. table \ref{notation} for a summary of the notation used in this paper. Figure adapted from Ref.~\cite{Schmekel:2018wbl}.}
\label{BYsetup}
\end{figure}

\begin{eqnarray}
E_1(t_f) - E_1(t_i) =
&&
-\int_{^3B} d^3x \sqrt{-\gamma}\, \tau_1^{ij} {\cal D}_i u_j
\nonumber\\
&&
+ \int_{^3B} d^3x \sqrt{-\gamma} u_\mu n_\nu T^{\mu\nu}.
\label{ConservationLaw}
\end{eqnarray}

Fig.~\ref{BYsetup} illustrates the geometric setup underlying this
balance law, showing the boundary $B$, its time evolution ${}^3B$, and
the associated unit normal vectors $n^\mu$ and $u^\mu$.

The second term represents ordinary stress-energy flux through the
boundary. The first term on the right hand side is purely geometrical. It can be written in a flux form, i.e. \cite{Schmekel:2018bcf}
\begin{eqnarray} \nonumber
& - & \int_{^3 B} d^3 x \sqrt{-\gamma} \tau_1^{ij} \mathcal{D}_i u_j = \\ \nonumber
& + &  \frac{1}{\kappa} \int_{^3 B} d^3 x \sqrt{-\gamma}  \nabla_k  \left [ \left ( \mathcal{D}_i u_j \right ) \left ( \gamma^{ij} \gamma^{kl} - \gamma^{ik} \gamma^{jl} \right ) \right ] n_l  \\ 
\label{FieldEnergyTerm}
\end{eqnarray}
An interpretation of the conservation of quasilocal momentum and angular momentum in terms of quasilocal quantities can be found in \cite{Epp:2013hua}.


\section{FLRW Cosmology}

We consider

\begin{equation}
ds^2 = -dt^2 + a^2(t) \left[ \frac{dr^2}{1-k_c r^2} + r^2 d\Omega^2 \right].
\end{equation}
where $k_c$ is either smaller than zero, zero or larger than zero depending on whether the universe
is open, flat or closed, respectively. $a(t)$ is the usual scale factor. 
Since the metric depends only on $a^2(t)$, the sign of $a(t)$ is a matter of convention rather than a
physical degree of freedom. We fix it assuming $a(t)>0$ which is consistent with the orientation of the
outward-pointing normal vector in eqn. \ref{unitnormals}.
For a comoving sphere,

\begin{equation}
u^\mu=\delta^\mu_t,
\qquad
n^\mu = \frac{\sqrt{1-k_c r^2}}{a} \delta^\mu_r .
\label{unitnormals}
\end{equation}

The nonvanishing components of the extrinsic curvature are

\begin{equation}
\Theta^\theta_{\ \theta} = \Theta^\phi_{\ \phi} = -\frac{\sqrt{1-k_c r^2}}{a r}.
\end{equation}

The Brown--York energy becomes \cite{Afshar_2009,Oltean:2020dvx,Chen:2007gx}

\begin{equation}
E_1 = -a r \sqrt{1-k_c r^2}.
\label{bareenergy}
\end{equation}

The quantity above represents the unreferenced Brown--York energy and should
not yet be interpreted as the physical matter energy enclosed by the boundary.


\section{Vanishing Matter Flux}

For a perfect fluid,

\begin{equation}
T_{\mu\nu} = (\rho+p)u_\mu u_\nu + p g_{\mu\nu},
\end{equation}

one immediately obtains

\begin{equation}
u_\mu n_\nu T^{\mu\nu}=0.
\label{AbsenceMatterFlux}
\end{equation}

Thus, no ordinary stress-energy crosses the comoving boundary.
This result is independent of the equation of state and therefore holds
for dust, radiation and vacuum energy alike.

Nevertheless, $\dot E_1 \neq 0$. The Brown--York balance law therefore implies

\begin{equation}
\dot E_1 = - \int_B d^2x \sqrt{\sigma} \tau_1^{ij} {\cal D}_i u_j .
\end{equation}

Hence, the entire evolution of the Brown--York energy is carried by a
purely geometrical boundary term which can be interpreted as a flux of
energy through the boundary as indicated by eqn. \ref{FieldEnergyTerm}. 
At this stage, however, this term should not yet be identified with the energy loss associated with
cosmological redshift, since such an interpretation requires a physically meaningful reference subtraction.
As an explicit check on the balance law, we now verify directly that
integrating the geometric flux term reproduces the value of
$E_1(t_f)-E_1(t_i)$ already obtained from eqn. \ref{bareenergy}.

\begin{eqnarray}
E_1(t_f) - E_1(t_i) = - \left . a(t) \cdot r \sqrt{1-k_c r^2} \right |^{t_f}_{t_i}
\end{eqnarray}
where
\begin{eqnarray}
\tau_1^{ij} \mathcal{D}_i u_j = \frac{1}{4 \pi} \frac{\dot a(t)}{a^2(t)} r^{-1}  \sqrt{1-k_c r^2}
\end{eqnarray}

The integral can be evaluated with $\sqrt{-\gamma} = r^2 a^2(t) \sin \theta$,
and one finds that it exactly reproduces $E_1(t_f)-E_1(t_i)$ as given
above. This explicitly confirms that the balance law, eqn.~\ref{ConservationLaw},
holds in this case, with the entire change in $E_1$ accounted for by
the geometric boundary term alone.

\section{Dynamical Reference Geometry}

Up to this point no reference subtraction has been employed.
The Brown--York balance law itself is completely independent of the
choice of reference action and therefore provides the natural starting
point of the present analysis.
Beginning from this point we restrict our treatment to flat universes with $k_c=0$.

In the original Brown--York formalism, the quasi-local energy is
defined only up to an additive reference term arising from an
additional boundary action $S_0=S_0[\gamma_{ij}]$,
where $\gamma_{ij}$ denotes the induced metric of the timelike
boundary world tube ${}^3B$ \cite{Brown:1992br}.
The formalism itself does not uniquely determine the explicit form of
$S_0$. Different choices therefore merely correspond to different
choices of the zero point of energy, while leaving the Einstein field
equations unchanged.

In many applications, the reference geometry is chosen by embedding
the two-dimensional boundary into Euclidean three-space. This
prescription successfully removes the leading divergence in
asymptotically flat spacetimes and reproduces the ADM mass at spatial
infinity. However, this construction is not uniquely required by the
Brown--York formalism itself.

For a comoving spherical boundary in FLRW spacetime, the induced
metric on the timelike world tube is

\begin{equation}
ds_{{}^3B}^2 = -d\tau^2 + R^2(\tau) d\Omega^2 ,
\label{worldtubemetric}
\end{equation}

where

\begin{equation}
R(\tau)=a(\tau)r
\end{equation}

is the areal radius.

Unlike the static situations usually considered in the literature,
the intrinsic geometry of the boundary itself is therefore explicitly
time dependent. Since the reference action fundamentally depends on
the full boundary metric $\gamma_{ij}$ rather than on an individual
spatial cross section alone, it is natural to consider a reference
geometry which reproduces the complete dynamical geometry of the
world tube.

A spherical world tube with radius $R(\tau)$ may be embedded into
Minkowski spacetime according to

\begin{equation}
X^\mu = \left( T(\tau), R(\tau), \theta, \phi \right),
\end{equation}

with induced metric

\begin{equation}
ds^2 = \left( -\dot T^2+\dot R^2 \right) d \tau^2  + R^2(\tau) d\Omega^2 .
\end{equation}

Comparison with eqn. \ref{worldtubemetric} immediately yields

\begin{equation}
-\dot T^2+\dot R^2=-1,
\end{equation}

which, choosing the branch with $\dot T>0$ so that $X^\mu(\tau)$
remains future-directed, yields

\begin{equation}
\dot T = \sqrt{1+\dot R^2}.
\label{Gamma}
\end{equation}

For a comoving sphere in FLRW spacetime one has

\begin{equation}
\dot R = HR,
\end{equation}

with the Hubble parameter $H \equiv \dot a / a$.   
The outward-pointing unit normal to the spherical worldtube in the Minkowski reference spacetime is $n_0^\mu=(\dot R,\dot T,0,0)$,
which satisfies $u_{0\mu}n_0^\mu=0$ and $n_{0\mu}n_0^\mu=1$. The trace of the extrinsic curvature of a spherical cross-section with respect to this normal is therefore
$k_0= - \sigma^A_B \nabla_A n_0^B = - 2n_0^r / R = - 2\dot T / R$. 
Consequently, the corresponding reference contribution to the Brown-York surface energy is
\begin{eqnarray}
E_0 = \frac{1}{8\pi}\int_B d^2x \sqrt{\sigma} k_0 = - R\dot T.
\end{eqnarray}

Using eqn. \ref{Gamma}, together with $\dot R=HR$ for a comoving sphere, gives
\begin{eqnarray}
E_0 = - R\sqrt{1+H^2R^2}.
\label{DynamicReference}
\end{eqnarray}

The corresponding reference-subtracted Brown--York energy is

\begin{equation}
E_{\rm BY} = E_1 - E_0 = R \left( \sqrt{1+H^2R^2} - 1 \right).
\label{ReferenceSubtracted}
\end{equation}

The present construction differs from the original Euclidean
embedding prescription employed by Brown and York.
However, the formalism itself only requires the existence of a
functional $S_0=S_0[\gamma_{ij}]$ eqn. (\ref{S0})
and does not uniquely specify its explicit form.
The dynamical reference considered here therefore represents an
alternative but fully admissible choice of reference action adapted
to the intrinsic geometry of the timelike boundary world tube.

\section{A Dynamical Criterion for the Choice of the Reference Term}
\label{sec:reference}

The Brown--York reference action $S_0[\gamma_{ij}]$ is only required to
depend on the intrinsic metric of the timelike boundary.
Consequently, the associated reference stress tensor

\begin{equation}
\tau^{ij}_0 = \frac{2}{\sqrt{-\gamma}} \frac{\delta S_0}{\delta\gamma_{ij}}
\label{tau0}
\end{equation}

is not uniquely determined.
Traditionally, this freedom has been interpreted primarily as a freedom
to choose the zero point of the quasi-local energy.
However, the Brown--York balance law suggests a considerably stronger
interpretation.

The quasi-local energy on a spatial section $B$ is determined by the
projection

\begin{equation}
\varepsilon = u_i u_j \tau^{ij},
\end{equation}

where $\tau^{ij}=\tau_1^{ij} - \tau_0^{ij}$, 
so that the instantaneous energy depends only on the ''$tt$-component'' of
the boundary stress tensor.
Its evolution, however, is governed by the integrated balance law,

\begin{equation}
\Delta E = \int_{{}^3B}  d^3x  \sqrt{-\gamma}  T_{\mu\nu}n^\mu u^\nu - \int_{{}^3B} d^3x \sqrt{-\gamma} \tau^{ij} {\cal D}_i u_j .
\end{equation}

The second term involves the complete boundary stress tensor rather
than only its energy density.
Consequently, the reference contribution enters through

\begin{equation}
\int_{{}^3B} d^3x \sqrt{-\gamma} \tau^{ij}_0 {\cal D}_i u_j ,
\end{equation}

which generally depends on the spatial stresses and momentum densities
of the reference tensor.

This observation has an important conceptual consequence.
A reference functional does not merely define the zero point of the
quasi-local energy itself.
It also defines the zero point of its dynamical evolution.
Unlike adding a constant energy offset, $E \rightarrow E+C$,
which leaves energy differences invariant, a dynamical reference functional changes
$\Delta E \rightarrow \Delta E - \Delta E_0$, where
\begin{equation}
\Delta E_0 = - \int_{{}^3B} d^3x \sqrt{-\gamma}  \tau^{ij}_0 {\cal D}_i u_j .
\end{equation}

Thus, different admissible reference functionals may lead to different
predictions for the evolution of the quasi-local energy, even when the
instantaneous energy on each spatial section is normalized in the same
way.

The spatially flat FLRW spacetime provides a particularly transparent
example.
Using the conventional instantaneous subtraction based on a static
Euclidean embedding yields  \cite{Afshar_2009}

\begin{equation}
E_{\rm BY}=0
\end{equation}

at every instant.
Consequently,

\begin{equation}
\Delta E_{\rm BY}=0,
\end{equation}

although the cosmological matter content evolves.
The vanishing result follows because the evolution of the unreferenced
Brown--York energy is exactly cancelled by the work performed by the
reference stress tensor.

This suggests that the choice of reference should not only reproduce an
appropriate energy normalization but should also preserve physically
meaningful dynamical information.

An independent consistency condition is provided by the small-sphere limit.
Brown, Lau, and York \cite{Brown:1998bt} showed that for a regular spacetime containing
matter the quasi-local energy evaluated using a light-cone reference construction - 
a specific isometric embedding of the shrinking two-surface into a genuine light cone of
Minkowski spacetime, rather than an arbitrary choice of reference - satisfies

\begin{equation}
E(R) = \frac{4}{3} \pi R^3 u^\mu u^\nu T_{\mu\nu}  + \mathcal{O} (R^4),
\label{smallsphere}
\end{equation}

while the leading contribution in vacuum occurs only at order $R^5$.
Crucially, this result is not automatic for an arbitrary reference
functional: as shown above, the conventional static Euclidean
embedding applied to a spatially flat FLRW universe yields
$E_{\rm BY}=0$ identically, in manifest violation of
eqn. \ref{smallsphere}. The small-sphere matter-recovery property is
therefore a genuine, discriminating criterion on $S_0$ - satisfied
by the light-cone reference of Brown, Lau, and York, by Epp's
invariant construction \cite{Epp:2000}, and by the dynamical
worldtube reference introduced here, but not by the static Euclidean
embedding - rather than a feature shared automatically by every
admissible choice of reference.

This expresses the physically expected result that sufficiently small
regions contain only local stress-energy to leading order, whereas
gravitational energy is intrinsically non-local and appears only in
higher orders.

The small-sphere limit therefore constrains the admissible reference
energy density.
The balance law discussed above suggests an additional requirement:
the associated reference stress tensor should not remove genuine
physical evolution through the reference work term.
From this perspective, the choice of $S_0$ is not merely a choice of
energy normalization but a choice of the complete dynamical reference
stress tensor on the timelike boundary.

The worldtube reference introduced in the present work naturally
satisfies this viewpoint.
Rather than defining the reference independently on each spatial
section, the complete timelike boundary is embedded into Minkowski
spacetime.
The resulting reference action therefore determines the full
three-dimensional reference stress tensor and its contribution to the
Brown--York balance law in a geometrically consistent manner.

Whether this dynamical criterion uniquely selects a preferred reference
functional remains an open question.
Nevertheless, it appears to provide additional physical guidance beyond
the traditional interpretation of the reference term as merely fixing
the zero point of the quasi-local energy.

\subsection{Desirable Properties of the Reference Functional}

For the specific purpose of the present work, namely the separation of
matter and gravitational contributions to leading order in the
small-sphere expansion, the following properties provide useful
criteria for the choice of reference functional. They are not intended
to define a unique or universally preferred reference prescription;
different choices may emphasize different physical aspects of the
quasi-local geometry.

In particular, any reference intended to identify the leading-order
quasi-local energy with the enclosed matter energy must reproduce the
matter small-sphere limit. This requirement is therefore not imposed
as a universal criterion on all Brown-York reference choices, but only
for the specific matter-field separation considered here.

\begin{enumerate}

\item
The small-sphere limit should reproduce the local matter energy (cf. \ref{smallsphere})
while purely gravitational contributions should first appear at higher
order.

\item
The reference stress tensor should provide a physically meaningful
contribution to the Brown--York balance law,

\begin{eqnarray}
\Delta E & = &   \int_{^3 B} d^3 x \sqrt{-\gamma} n^\mu u^\nu T_{\mu\nu} 
\nonumber \\
& - & \int_{^3 B} d^3 x \sqrt{- \gamma} ( \tau_1^{ij}-\tau^{ij}_0 ) {\cal D}_i u_j,
\label{ConservationLawReferenced}
\end{eqnarray}

without removing genuine physical evolution through the reference work
term.

\item
The reference should be defined consistently on the entire timelike
boundary rather than independently on each spatial section whenever
dynamical evolution is considered.

\end{enumerate}

It is worth emphasizing that property 1 does not single out a unique
$S_0$. Rather, it identifies an equivalence class of admissible
reference functionals --- including the light-cone reference of
Brown, Lau, and York \cite{Brown:1998bt}, Epp's invariant construction
\cite{Epp:2000}, and the dynamical worldtube reference introduced
here - all of which agree at leading order in the small-sphere
expansion while differing at higher orders. 

\subsection{Relation to Other Quasi-Local Constructions}

The use of Minkowski spacetime as a reference geometry has several
precursors. In the original Liu--Yau construction \cite{PhysRevLett.90.231102}, the physical
surface term is formulated in terms of the spacetime mean-curvature
vector, while the reference contribution is still obtained from an
isometric embedding of the two-surface into Euclidean three-space.
A subsequent generalization by Wang and Yau \cite{WangYau2007} employed an isometric
embedding into a hyperboloid in Minkowski spacetime and produced a
future-directed timelike quasi-local energy-momentum vector. The later
Wang--Yau construction \cite{PhysRevLett.102.021101} then used general isometric embeddings of the
two-surface into Minkowski spacetime together with a choice of reference
observer.

Epp's invariant quasi-local energy, for example,
is based on an isometric embedding of the two-surface into a
four-dimensional reference spacetime rather than on an embedding into
a three-dimensional spatial reference slice
\cite{Epp:2000}. Related boost-invariant constructions include the
Liu--Yau quasi-local mass \cite{PhysRevLett.90.231102}.

The present construction is conceptually closer to earlier
Minkowski-reference approaches such as the generalized
Liu--Yau construction of Wang and Yau than to the later
optimal embedding formulation by Wang--Yau.

There is nevertheless an important conceptual difference. In these
constructions the primary object is the spacelike two-surface and its
normal bundle. The present construction differs from all of these prescriptions in
that the reference object is not an individual spacelike two-surface.
In the present approach, by contrast, the primary
reference object is the complete timelike boundary world tube
${}^3B$. 

Boost-invariant quasi-local energies applied to spherical cosmological
boundaries frequently lead to expressions of the form \cite{Afshar_2009}

\begin{equation}
E_{\rm Epp} = R \left[ 1- \sqrt{ 1- H^2 R^2} \right],
\label{InvariantQLE}
\end{equation}

up to conventions and the precise choice of observers and reference
term. The dynamical world-tube reference considered here is given 
by eqn. \ref{ReferenceSubtracted} instead.
The two expressions should therefore not be identified. In
eqn. \ref{InvariantQLE} the cosmological dependence enters through
the norm of the mean-curvature vector of a two-surface. In
eqn. \ref{ReferenceSubtracted} it originates from the Lorentz factor
required to embed the complete timelike world tube into the reference
spacetime.

Nevertheless, both constructions possess the same leading
small-sphere behavior,

\begin{equation}
E = \frac12 H^2R^3 + {\cal O}(R^5) = \frac{4}{3} \pi \rho R^3 + {\cal O}(R^5),
\end{equation}

for a spatially flat FLRW universe satisfying the Friedmann equation.
The agreement of the leading term is therefore not sufficient to
identify the two prescriptions, but it demonstrates that both recover
the expected matter contribution in the small-sphere limit.

\section{Small-Sphere Expansion}

The reference-subtracted Brown--York energy,
eqn. \ref{ReferenceSubtracted}, admits the expansion

\begin{equation}
E_{\rm BY} = \frac12 H^2R^3 - \frac18 H^4R^5 + {\cal O}(R^7).
\label{SmallSphere}
\end{equation}

Using the Friedmann equation

\begin{equation}
H^2 = \frac{8\pi}{3}\rho,
\label{Friedmann}
\end{equation}

the leading contribution becomes

\begin{equation}
E_{\rm BY} = \frac{4}{3} \pi \rho R^3 - \frac18 H^4R^5 + {\cal O}(R^7).
\label{MatterLimit}
\end{equation}

The first term coincides precisely with the matter energy enclosed
inside the sphere (cf. \cite{Brown:1998bt,Schmekel:2018bcf}),

\begin{equation}
E_{\rm matter} = \rho V = \frac{4}{3} \pi \rho R^3.
\end{equation}

Thus,

\begin{equation}
E_{\rm BY} = E_{\rm matter} + {\cal O}(R^5),
\end{equation}

showing that the dynamical reference subtraction reproduces the
expected small-sphere matter limit.

The next contribution may be written as

\begin{equation}
E_{\rm self} = -\frac18 H^4R^5 .
\end{equation}
Using eqn. \ref{Friedmann} together with

\begin{equation}
E_{\rm matter} = \frac12 H^2R^3 ,
\end{equation}

one obtains

\begin{equation}
E_{\rm self} = - \frac{E_{\rm matter}^2}{2R}.
\label{SelfEnergy}
\end{equation}

For comparison, the Newtonian self-energy of a uniform-density sphere
of mass $M$ and radius $R$, obtained by assembling the sphere from
concentric shells, is
\begin{equation}
E_{\rm Newton} = -\int_0^R \frac{m(r)}{r}\,dm = -\frac{3}{5}\frac{M^2}{R},
\end{equation}
where $m(r)=M(r/R)^3$ for uniform density and $dm=4 \pi \rho r^2 dr$. 
This yields $E_{\rm Newton}=-\tfrac{3}{5}M^2/R$, which
coincides in functional form --- though not in numerical prefactor ---
with eqn. \ref{SelfEnergy}. Since the FLRW matter density is exactly
uniform within the small sphere considered here, this comparison
applies directly at leading order, with $M\to E_{\rm matter}$. The
$\sim20\%$ discrepancy in the prefactor ($1/2$ vs.\ $3/5$) is not
unexpected, since $E_{\rm self}$ is a relativistic quasi-local
quantity and need not coincide exactly with its Newtonian counterpart
even at leading nontrivial order; the essential point is the shared
$M^2/R$ scaling and negative sign characteristic of gravitational
binding energy.
The higher-order terms in the Brown--York expansion therefore admit
a natural interpretation as nonlinear gravitational corrections to
the leading matter contribution.

\section{Radiation and Cosmological Redshift}

For radiation,

\begin{equation}
p=\frac{\rho}{3},
\end{equation}

the continuity equation yields

\begin{equation}
\dot \rho + 4H \rho = 0,
\end{equation}

and therefore

\begin{equation}
\rho \propto a^{-4}.
\end{equation}

The matter energy enclosed inside a comoving sphere becomes

\begin{equation}
E_{\rm matter} = \frac{4}{3} \pi R^3 \rho \propto a^{-1}.
\label{PhotonEnergy}
\end{equation}

Equation~(\ref{PhotonEnergy}) represents the standard cosmological
redshift relation.

Consequently, the leading contribution to the quasi-local energy
balance reproduces precisely the energy decrease associated with
cosmological redshift.
At higher orders, additional nonlinear gravitational contributions
appear naturally through eqn. \ref{SelfEnergy}.
These terms may be interpreted as gravitational self-energy
corrections associated with the finite quasi-local system.

The present analysis therefore suggests that the apparent loss of
photon energy does not necessarily indicate a breakdown of energy
conservation. Rather, cosmological redshift may be understood as part of the
quasi-local energy balance (cf. eqns. \ref{ConservationLaw}, \ref{ConservationLawReferenced})
of the combined matter-gravity system.

The absence of ordinary matter flux through the comoving boundary eqn. \ref{AbsenceMatterFlux}, i.e. $u_\mu n_\nu T^{\mu\nu}=0$,
further emphasizes that this energy accounting is intrinsically
quasi-local and cannot be interpreted as a local transfer of energy
across the boundary itself.

\section{Discussion}

The present results suggest that the apparent energy problem
associated with cosmological redshift does not require abandoning
the notion of energy conservation, but rather calls for a
quasi-local rather than local or global formulation thereof.
The Brown--York balance law provides an exact quasi-local
conservation law in which the evolution of the enclosed energy is
governed entirely by boundary quantities.

For comoving observers in FLRW spacetime, ordinary matter fluxes
through the boundary vanish identically.
The evolution of the unreferenced Brown--York energy is therefore
entirely determined by geometrical boundary contributions, which
admit a direct mechanical interpretation as work performed by the
Brown--York surface stresses.

The introduction of a dynamical reference geometry adapted to the
intrinsic geometry of the timelike boundary world tube leads to a
remarkably simple small-sphere expansion.
Its leading contribution reproduces exactly the matter energy
contained within the boundary, while higher-order terms naturally
possess the structure expected for gravitational self-energy
corrections.

In particular, for radiation dominated cosmologies the leading
quasi-local contribution reproduces the standard cosmological
redshift relation. This suggests that the apparent loss of photon
energy may be understood within a fully quasi-local conservation
framework.

It is important to emphasize that this conclusion does not rest on
the particular choice of dynamical reference introduced here. As
discussed in Sec.~VI.A, any reference functional satisfying the
small-sphere matter-recovery property - including the light-cone
reference of Brown, Lau, and York, Epp's invariant construction, and
the worldtube reference of the present work - agrees at leading
order in the small-sphere expansion, even though these constructions
differ at higher orders and are motivated by different physical
considerations. Since the present analysis requires only a separation
of matter and field energy to leading order, the identification of
the redshift-associated energy loss as a purely geometric boundary
flux is therefore a robust feature of this equivalence class of
references, rather than an artifact of one specific choice of $S_0$.

\section{Conclusions}

The Brown-York formalism admits considerable freedom in the choice of the reference action $S_0$. We do not attempt to characterize this freedom in general. Instead, we argue that physically motivated criteria should guide the choice of reference geometry.

In particular, a physically meaningful reference should reproduce the correct small-sphere limit while remaining consistent with the Brown--York conservation law. The dynamical reference considered in this work satisfies these requirements and provides a natural quasi-local interpretation of cosmological redshift.

Within this framework, the apparent loss of photon energy in an expanding radiation-dominated universe is not associated with a flux of matter across the boundary. Instead, it is entirely accounted for by the geometrical boundary flux appearing in the Brown--York conservation law. 
Consequently, the interpretation presented here should not be
regarded as a consequence of one particular choice of reference
action. Rather, because only a leading-order separation of matter and
field energy is required for this purpose, it is a generic feature
shared by any admissible reference satisfying this minimal criterion -
a requirement met, as shown in section VI.A, by several
independently motivated constructions, though not by the conventional
static Euclidean embedding.
Whether these physical requirements uniquely determine the reference action remains an interesting mathematical question, but lies beyond the scope of the present work.

\acknowledgments
The author would like to acknowledge insightful discussions with James W. York, Jr., Andrew P. Lundgren and Shing-Tung Yau.

\bibliography{bib}
\bibliographystyle{h-physrev}

\appendix
\section{Intermediate Results}

We list the non-vanishing components of the extrinsic curvature and the stress-energy-momentum surface density tensor
\begin{eqnarray}
\Theta^\theta_\theta=\Theta^\phi_\phi=-\frac{\sqrt{1-k_c r^2}}{r a(t) } \\
\tau^{\theta \theta} = \frac{\sqrt{1-k_c r^2}}{\kappa r^3  a^3(t) } \\
\tau^{\phi \phi} = \frac{\sqrt{1-k_c r^2}}{\kappa r^3 a^3(t)  \sin^2 \theta} \\
\tau^{t t} = -\frac{2 \sqrt{1-k_c r^2}}{\kappa r a(t) }
\end{eqnarray}

\section{Notation}

\begin{widetext}

\begin{table}
\begin{tabular}{|r|c|c|c|c|c|} \hline
Manifold	& (Induced)	&	Covariant 	&	Unit normal 	&	Intrinsic	&	Extrinsic  \\ 
	&	metric	&	derivative	&	vector		&	curvature	&	curvature \\ \hline
$M$		&	$g_{\mu \nu}$ &	$\nabla_{\mu}$ &	&	$\mathcal{R}_{\mu \nu \rho \sigma}$ &	\\
$\Sigma$		&	$h_{ij}$	&	$D_i$	& $u_{\mu}$	&	$R_{ijkl}$ & $K_{ij}$ \\
	&	$h_{\mu \nu}=g_{\mu \nu}+u_\mu u_\nu$ & $D_\mu t^\nu = h^\alpha_\mu h^\nu_\beta \nabla_\alpha t^\beta$ & & & $K_{\mu \nu} = - h_{\mu}^{\lambda} \nabla_{\lambda} u_{\nu}$ \\ 
$^3 B$ 		&	$\gamma_{\mu \nu}$ & $\mathcal{D}_{i}$ &	$n_{\mu}$		&	&	$\Theta_{\mu \nu}$ \\
			&	$\gamma_{\mu \nu}=g_{\mu \nu}-n_\mu n_\nu$ &   $\mathcal{D}_\mu t^\nu = \gamma^\alpha_\mu \gamma^\nu_\beta \nabla_\alpha t^\beta$ & &  & $\Theta_{\mu \nu}=-\gamma_{\mu}^{\lambda} \nabla_{\lambda} n_{\nu}$ \\
$B$		&	$\sigma_{\mu \nu}$	&	&	$n_{\mu}$ &	&	$k_{\mu \nu}$ \\
		&	$\sigma_{\mu \nu} =g_{\mu \nu} - n_\mu n_\nu + u_\mu u_\nu$ & & & & $k_{\mu \nu}=-\sigma_{\mu}^{\lambda} D_{\lambda} n_{\nu}$ \\
\hline
\end{tabular}
\caption{Summary of notation. }
\label{notation}
\end{table}

\end{widetext} 

\end{document}